\documentclass{llncs}

\newcommand{\chr}{{CHR}}

\def\example{\par\medskip\noindent\textbf{Example:\quad}}

\begin{document}

\title{CHR as grammar formalism\\
{\it A first report}}
\author{Henning Christiansen}
\institute{Roskilde University,
Computer Science Dept.,\\
P.O.Box 260, DK-4000 Roskilde, Denmark\\
E-mail: \email{henning@ruc.dk}}%%%%%\\

\maketitle

\begin{abstract}
Grammars written as Constraint Handling Rules (CHR)
can be executed as efficient
and robust bottom-up parsers that provide a
straightforward, non-backtracking treatment of ambiguity.
Abduction with integrity constraints as well
as other dynamic hypothesis generation techniques
fit naturally into such grammars
and are exemplified for anaphora resolution, coordination
and text interpretation.
\end{abstract}

\section{Introduction}
The language of Constraint Handling Rules~\cite{fruehwirth-95,fruehwirth-98},
CHR, was introduced as a
tool for writing constraint solvers for traditional constraint domains
such as real or integer number arithmetic and finite domains.
CHR, however, has turned out to be of more general value
and is
now available as an extension of among others,
SICStus  Prolog~\cite{sicstus-manual}.
The CHR web pages~\cite{chronline} contains a growing
collection of applications.
Being of special interest to language processing,
\cite{AbdennadherSchuetz98} has shown that CHR adds
bottom-up evaluation to Prolog and a flexibility
to combine top-down and bottom-up;
\cite{AbdChr2000} has taken this a step further showing that
abduction and integrity constraints can be expressed in
a straightforward way in CHR.

Here we investigate CHR applied as a grammatical formalism
and it appears that a grammar can be written in a
straightforward way as propagation rules of CHR that execute
as a robust bottom-op parser which handles ambiguity
without backtracking.
Constructed in the most naive way,
these parsers are of a high time-complexity, typically
cubic or worse, but by means
the so-called pragmas provided by current implementations
of CHR, realistic and almost linear parsers are achieved,
executing quite similary to
shift-reduce parsers.

Look-ahead techniques are easily expressed and
ambiguous grammars can be made unambiguous using simplification
rules instead of propagation rules.
Similarly as for 
Definite Clause Grammars~\cite{PereiraWarren1980}
(DCGs), CHR grammars
(as we deliberately call them)
can take arbitrary attributes
in a similar way but accepts a wider class of context-free
bases. Except for direct loops in the grammar, CHR grammars works
correctly for any such basis.
As is well-known, DCG's top-down evaluation with backtracking may 
lead to combinatorial explosion in time which is avoided
by a bottom-up evaluation that examines all hypotheses
in the same space.
However, there are other and more significant differences,
most importantly the way in which
arbitrary hypotheses can be produced and consumed in CHR
and applied for
controlling the parser and for evaluation of semantics.
Assumption grammars~\cite{DahlTarauLi97} that use
linear and intuitionistic assumptions in
various ways can be implemented very easily
in CHR and abduction with integrity constraints
works fine as well.
Examples are given of anaphora resolution, coordination, and text
interpretation by abduction.

\subsection*{Related work}
The notion of constraints, with slightly
different meanings, is often
associated with language processing.
``Constraint grammars'' and ``unification grammars''
is often used
for feature structure grammars
and constraint programming techniques have been applied
for the complex constraints that arise in natural
language processing, see, e.g.,~\cite{Allen1995,Duchier2000}
for introduction and overview.
Constraint logic programming, in terms of using black-box
constraint solvers, has been applied at many occasions
as well.

What we propose is to
capture fundamental processes such as parsing
and semantic analysis by means of declaratively
specified constraint solvers, more specifically,
expressed and implemented in the language of CHR.
While our results are promising, there is still very little
work to report in this context.
The basic principle of using CHR's propagation rules
for context-free
parsing has also been suggested by~\cite{Meyer2000}
but not elaborated further;
\cite{Penn2000} has used CHR for evaluation of feature
structures for HPSG.

Assumption Grammars~\cite{DahlTarauLi97},
inspired by Linear Logic, provide a structured way to
generate and manage dynamically generated hypothesis.
This provides elegant solutions to problems
that are otherwise difficult to handle, although
there seems to be a need for more detailed methods
to control the scope of such hypotheses.
We do not provide new suggestion for this, but
it seems possible to implement a variety of such
mechanisms in our CHR based framework.

The close similarity between our representation of
abduction and of Assumption Grammars in CHR
seems to indicate
a closer relation between the two that still needs
to be investigated.
References related to abduction
are given in section~\ref{abduction-section}.

\subsection*{Overview}
Section~\ref{basics-section} explains how propagation rules
provide obviously correct parsers and
section~\ref{complexity-section} concerns their time complexity.
Section~\ref{assumption-grammar-section}
shows how assumption grammars can be implemented and discusses
some improvements and section~\ref{abduction-section}
introduces abduction and integrity constraints into the model.
The final section~\ref{conclusion-section} provides a
summary and discusses possible syntactic sugar
for CHR grammars and other perspectives.
No introduction to CHR is given; we refer to~\cite{fruehwirth-95,fruehwirth-98}
or online manuals~\cite{chronline,sicstus-manual}.

\section{CHR as grammar formalism: The basic
principle}\label{basics-section}
A set $P$ of propagation rules of \chr\ constitutes a natural
bottom-up evaluator: The set of initial hypotheses $H$
is given as a query and the evaluator produces as answer
the
set $A$ of all atoms that are logical consequences
of $P\cup H$, of course provided that no infinite loop
occurs. The set $H$ is the initial constraint store,
$A$ the final constraint store in which all rules of
$P$ that can apply have been applied.

Using \chr\ for parsing, we represent a string to
be analyzed as a set of initial constraints
of the form \texttt{token($t$,$n$$-$$1$,$n$)},
where $t$ is a Prolog atom and $n-1$, $n$ indicates
the position in the string. So, e.g.,
``Peter likes Mary'' is represented as
\begin{verbatim}
     token(peter,0,1), token(likes,1,2), token(mary,2,3).
\end{verbatim}
Nonterminals, say $Q$, of a context-free grammar can be represented as
a binary constraint symbol \texttt{$Q$($n$,$m$)} with the
interpretation ``substring between positions $n$ and $m$ comprises
an instance of $Q$''.
Each context-free grammar rule
\begin{flushleft}
     \texttt{\ \ \ \ }$Q_0 ::= Q_1,\ldots,Q_k$,
\end{flushleft}
with
$k>0$, $Q_0$ a nonterminal and each $Q_i$, $0\leq i\leq k$,
a terminal or nonterminal,
is represented as a
propagation rule
\begin{flushleft}
     \texttt{\ \ \ \ $S_1$,\ldots,$S_k$ ==> $Q_0$(N0,N$k$)}.
\end{flushleft}
where, if $Q_i$ is a terminal, $S_i=$ \texttt{token($Q_i$,$k$$-$$1$,$k$)},
and if $Q_i$ is a nonterminal, $S_i=$  \texttt{$Q_i$(N$i$$-$$1$,N$i$)}.

Empty productions $N::=\epsilon$ can be handled by inserting
\texttt{$N$(0,0)}, \texttt{$N$(1,1)}, etc.\ in the initial constraint store
but in the following we consider only grammars without empty productions.
We may distinguish a particular nonterminal of a grammar
as its start symbol.

\example The grammar
$\{$\textit{sentence}::= \textit{np verb np};
\textit{np}::= \textit{peter} $|$ \textit{mary};
\textit{verb::}= \textit{likes}$\}$ is translated
into the following CHR program.
\begin{verbatim}
     np(N0,N1), verb(N1,N2), np(N2,N3) ==> sentence(N0,N3).
     token(peter,N0,N1) ==> np(N0,N1).
     token(mary, N0,N1) ==> np(N0,N1).
     token(likes,N0,N1) ==> verb(N0,N1).
\end{verbatim}
A parser given in this way for a context-free grammar $G$
is obviously correct in
that the following properties hold;
let $t_1\cdots t_x$ be the input string and $S$ the start symbol.
\begin{enumerate}
     \item\label{noloop}  If the grammar contains no loops,
     i.e., no nonterminal can derive itself, the parser is guaranteed to
     terminate. Given this premise, we have also:

     \item\label{non-con}  If and only if
     nonterminal $N$ can generate substring
     $t_i\cdots t_j$, the final constraint store contains
     the constraint \texttt{$N$($i$,$j$)}.

     \item\label{overallCorrect} The final constraint store contains 
\texttt{$S$(0,$x$)}
     if and only if $S$ can derive $t_1\cdots t_x$.

     \item\label{abiguity-prop}
     If the grammar is ambiguous and a substring $t_i\cdots t_j$
     corresponds to $n$
     \break% to avoid overfull box at the end og this item
     different parse trees, each with top node
     $N_1,\ldots,N_n$, the final constraint store contains
     exactly $n$ constraints with argument list \texttt{($i$,$j$)},
     namely \texttt{$N_1$($i$,$j$)}, \ldots, \texttt{$N_n$($i$,$j$)}.
\end{enumerate}
Properties \ref{noloop} and \ref{overallCorrect} express the
usual
notion of correctness for a parser.
Property \ref{non-con} means that the parser is robust
in the sense that if the entire string does not conform
with the grammar, it is still possible to extract those parts
that represent recognizable subphrases.
Property \ref{abiguity-prop} shows that
ambiguity is handled in a natural way without backtracking;
the constraints can be extended with an argument
representing a parse tree (or some semantic representation)
and all possible trees (or meanings) can be read out of the final state.
These properties are desirable especially for natural language
processing where the language usually is richer than any
grammatical formalization of it and where ambiguity is
an inherent property.

However, property \ref{non-con} shows also that the final constraint
store may become quite large as it contains all nonterminals
that has been possible to recognize, even in case they
do not contribute to the overall parse due to
``local ambiguity''.
Using such a parser for the grammar
$\{$\textit{as}::= \textit{a}; \textit{as}::= \textit{a} \textit{as}$\}$
for a string of length $n$ will result in a final state
with $n(n+3)/2$ constraints.

\par\medskip\noindent
Extra arguments can be added to the nonterminals
in the same way as in a DCG and string indices can be hidden
by means of syntactic sugar \textit{\`a la} DCG.
In the final section, we discuss possible syntactic sugaring,
but until then we stay with our goal, namely
to investigate the full CHR language
as a grammatical formalism.
The declarative semantics of CHR provides a semantics for our grammars
but we will also refer to the procedural semantics and implemented features
for various adjustments.

\section{Time complexity and what to do about it}
\label{complexity-section}
Time complexity for constraint solvers similar to CHR's
propagation rules has been studied by
\cite{mcallester2000}.
It is shown that a parser
for a context-free grammar in Chomsky Normal Form
(max.\ two grammar symbols on rhs)
runs in time $gn^3$ where $g$ is the no.\ of grammar rules
and $n$ the length of the input string.
This conforms with the classical result for
the Cocke-Younger-Kasami parsing algorithm
that works quite similarly to a parser comprised
by propagation rules;
see, e.g.,~\cite{Aho-Ullman1972} for background.
We made an empirical test by analyzing random
sequences of $a$s and $b$s with a propagation rule parser
for the grammar
$G=\{$\textit{S}::=  \textit{a A\hskip -2pt B}  $|$  \textit{B A};
\textit{A}::= \textit{ B B\hskip -2pt B}  $|$  \textit{a};
\textit{B}::=  \textit{A S}  $|$  \textit{b};
\textit{A\hskip -2pt B}::=  \textit{A B};
\textit{B\hskip -2pt B}::=  \textit{B B}$\}$.
This grammar has an extreme degree of local ambiguity,
resulting in an huge number of derived constraints.
Our test with SICStus Prolog's implementation of
CHR shows indeed a complexity of $n^3$
until
an exponential factor takes over around $n=30$, probably due
to a very active
garbage collector.
For $n=30$, the final store
contains in average more that 1500 constraints.
The complexity increases to higher powers of $n$ when more
grammar symbols
appear in each rule.

We believe that more interesting grammars will
produce far less final constraints, but still
the $n^3$ makes the method
unrealistic for any practical application.
Fortunately, there are several ways to reduce
time complexity that we illustrate for a parser of
arithmetic expressions
based on a straightforward, ambiguous grammar.
One of the rules is the following.
\begin{verbatim}
     exp(N0,N1), token(+,N1,N2), exp(N2,N3) ==> exp(N0,N3).
\end{verbatim}
Current CHR implementations (that do not recognize
the pattern of recurrence of the variables) will test
this rule twice whenever a new \texttt{exp} is created,
once to check if it matches the leftmost \texttt{exp}
in the rule and secondly for the rightmost.
In addition, the entering of  \texttt{token(+,$\cdots$)}
will also result in a check for applicability of the rule.

CHR's so-called \texttt{pragma}s can be applied
to force the parser to work more like a classical shift-reduce
parser. We do this by making \textit{each but the rightmost}
grammar symbol passive in each rule, e.g.,
\begin{verbatim}
     exp(N0,N1)#Id1, token(+,N1,N2)#Id2, exp(N2,N3) ==> exp(N0,N3),
        pragma passive(Id1), passive(Id2).
\end{verbatim}
This means that whenever a new \texttt{exp} constraint is created,
only one test is generated for this rule, namely for a
potential match with the rightmost \texttt{exp} at the lhs of this rule.
The advent of \texttt{token(+,$\cdots$)} will never trigger this rule.
% It is assumed that \texttt{token} constraint are entered in the
% order in which they occur in the input string, e.i., with
% increasing indices. % EVIDENT FROM CHAPTER 2
It can observed\footnote{To prove this, one needs to argue in terms
of a fairly detailed model of CHR's
execution model~\cite{Abdennadher97}.}
that the modified constraint solver
produces exactly the same constraints as the original one
(and even in the same order), the parts removed from the
computation process are a lot of tests for applicability of
rules anyhow destined to fail.
We have not made a formal analysis of the time complexity for
these modified parsers but tests for a variety
of grammars indicate a linear or almost linear behaviour.

Another source of inefficiency is the potentially large number of
constraints that pile up in the state;
fewer final constraints will also make it easier to interpret
the result of the parsing process.
One way to reduce this
is to apply CHR's reduction rules instead of propagation rules,
i.e., write ``\texttt{<=>}'' instead of ``\texttt{==>}''.
The effect is that the constraints matched on the lhs of the rule
are removed from the constraint store and
thus fewer parse trees will be recognized.
Thus reduction rules should only be used
for parts of a grammar known to be unambiguous.
Alternatively, it can be seen as a method to make an ambiguous
grammar unambiguous and  a mixture
of propagation and reduction rules can be used.
Some degree of ambiguity, however, can be handled
in a reduction rule
by means of backtracking
using disjunctions at the
rhs of rule, cf.~\cite{AbdennadherSchuetz98},
or by launching all different hypothesis into the constraint store
at the same time.

Since the position of tokens are explicit in the rules
we can also let  parsing depend on other symbols
than the sequence being reduced.
One example of this is to incorporate a look-ahead
based on LR-items
(see, e.g.,~\cite{AhoSethiUllman1986}) as in the
following grammar for arithmetic expressions.
Here we use CHR's so-called simpagation rules:
Constraints matched to the left of the backslash stay
in the constraint store and the other ones are removed;
the test in front of the vertical bar is the a guard.
We assume the text is followed by the terminal symbol
\texttt{eof} and the following rule says
that ``\texttt{exp + exp}'' is reduced provided the
next input token is the specified list.
\begin{verbatim}
     token(R, N3,N4) \ exp(N0,N1), token(+,N1,N2), exp(N2,N3)
        <=> member(R,[+,')',eof]) | exp(N0,N3).
\end{verbatim}
The \texttt{passive} pragmas can be added as described
provided that ``right-most symbol'' in the lhs of a rule
is identified by means of the indices so that, in the rule above,
the \texttt{token} in front of the backslash
is the non-\texttt{passive} one and thus the one
and only that can trigger the rule.

Together with the following rules
(\texttt{pragma}s understood), we have a grammar
for arithmetic expressions with traditional associativity and
precedence.
\begin{verbatim}
     token(R,N3,N4) \ exp(N0,N1), token(*,N1,N2), exp(N2,N3)
        <=> member(R,[*,+,')',eof]) | exp(N0,N3).
     token(R,N3,N4) \ exp(N0,N1), token(^,N1,N2), exp(N2,N3)
        <=> R \= ^ | exp(N0,N3).
     token('(',N0,N1), exp(N1,N2), token(')',N2,N3) <=> exp(N0,N3).
     token(Int,N0,N1) <=> integer(Int) | exp(N0,N1).
\end{verbatim}
As a result we achieve an efficient
parser in which the constraint store is used
effectively as a stack and the parser performs
exactly the same steps and comparisons as a traditional LR(1) parser.

We conclude that time complexity
is not a problem for CHR parsers and  that
flexibility is available for the grammar writer
to consider symbols (not only terminals!) to the left and to right of the
symbols reduced, to combine CHR's different kinds of rules and
pragmas with different effects, to add new tokens
on the rhs (not shown), etc., etc.
In addition, the parser may refer to constraints that
represent other than purely syntactic hypothesis;
this appears in the following.

\section{Assumption Grammars and beyond}\label{AGs}
\label{assumption-grammar-section}
With CHR it is possible to let rules produce and consume
arbitrary hypotheses.
Consider the following sketch of two CHR grammar rules,
\begin{verbatim}
         ...      ==> ... h(X) ... .
     ... h(X) ... ==>     ...      .
\end{verbatim}
The first rule produces a hypothesis, say \texttt{h(a)}, extracted
from the context in which the rule is applied and the second rule
can be executed when such a hypothesis is present and the value
\texttt{X=a} becomes available. Anaphora in natural language can
be treated by passing hypotheses through the constraint
store in this way.

Assumption grammars~\cite{DahlTarauLi97} include specialized operators
for managing such hypotheses and we show how they
can be implemented and applied in CHR grammars.
In an assumption grammar, the expression $+h(a)$ asserts a linear
hypothesis which can be used once in the following text by means
of the expression $-h(a)$ (or
$-h(X)$), called an expectation.
Asserting the hypothesis by $*h(a)$ means that it can be
used over and over again. We represent an assertion $+h(a)$ by a
constraint \texttt{+(h,[a],$n$)}; the predicate symbol
and argument list are split for technical reasons and $n$
indicates a position in the string where the
hypothesis is supposed to be created.
The two other operators are represented in analogous ways.
Deviating slightly
from the syntax of~\cite{DahlTarauLi97}
(as to achieve a more symmetric notation),
we introduce three new operators for ``time-less''
hypotheses, \texttt{=+}, \texttt{=-}, and \texttt{=*},
whose meanings are similar except that hypotheses can be
used and consumed in any order; for these we can leave out
the string index.
The following simplification and simpagation
rules provide an implementation in CHR.
\begin{verbatim}
     =+(P,A)   , =-(P,B)   <=>    true & A=B | true.
     =*(P,A)   \ =-(P,B)   <=>    true & A=B | true.
     +(P,A,Z1) , -(P,B,Z2) <=> Z1 < Z2 & A=B | true.
     *(P,A,Z1) \ -(P,B,Z2) <=> Z1 < Z2 & A=B | true.
\end{verbatim}
The explicit unification in the guard serves as test for unifiability
as well as for porting the argument value from the asserted
hypothesis (supposed to be ground) to its application.\footnote{CHR
matches constraints to be processed by a common ``instance of''
condition, so that the rule \texttt{=+(P,A), =-(P,A) <=> true} would
apply in fewer cases than the corresponding one above.}
Correct derivation in an assumption grammar requires that all expectations
are matched by corresponding assertions in the end
and an optional
test \verb!all_consumed! is available meaning that all linear hypotheses have
been consumed; these and similar facilities are easily
implemented by means of CHR's auxiliary predicates.\footnote{E.g.,
\texttt{all\char95 consumed:-
\char92+ find\char95 constraint(+(\char95,\char95,\char95),\char95),}\\
\texttt{\char92+ find\char95 constraint(=+(\char95,\char95,\char95),\char95)}.}

However, the use of hypotheses introduces one technical
problem in CHR,
because a give parse tree has its specific set of hypotheses.
Thus, with an ambiguous grammar and propagation rules
we would mix up hypotheses corresponding to different trees.
For the present we assume unambiguous grammars
and use simplification rules.
We give excerpts from a sample grammar inspired by~\cite{DahlTarauLi97}
demonstrating anaphora and coordination.
An occurrence of a proper name for
some individual (``\texttt{X}'' in the following rule)
introduces a hypothesis that this individual
can be referred to subsequently by a pronoun.
\begin{verbatim}
     proper_name(X,Gender,N1,N2)  <=>
        *(active_individual,[X,Gender],N1), np(X,Gender,N1,N2).
     pronoun(Gender,N1,N2) <=>
        -(active_individual,[X,Gender],N1), np(X,Gender,N1,N2).
\end{verbatim}
Coordination arise in relation to ellipses as
in ``Mary likes and Martha hates Peter''.
The object for the first sentence is implicit and shared
with the second sentence.
Simple and full sentences are described as follows.
\begin{verbatim}
     np(Sub,_,N1,N2), verb(V,N2,N3), np(Obj,_,N3,N4) <=>
        sent(V-(Sub,Obj),N1,N4).
\end{verbatim}
The following two rules take care of coordination.
Notice that the ellipsis is identified by a look-ahead
for ``and''; the second of two and'ed sentences offers its
object to anyone missing its object.
\begin{verbatim}
     token(and,N3,N4) \ np(Sub,_,N1,N2), verb(V,N2,N3) <=>
        =-(ref_object,[Obj]), sentence(V-(Sub,Obj),N1,N3).
     sent(S1,N1,N2), token(and,N2,N3), sent(V2-(Sub2,Obj2),N3,N4)
      <=> =+(ref_object,[Obj2]), sent(S1+(V2-(Sub2,Obj2)), N1,N4).
\end{verbatim}
With these and a few other obvious rules, the analysis of
``Mary likes Peter. She loves and Martha hates him.''\ yields
the following semantic representation.
\begin{verbatim}
     likes-(mary,peter) + loves-(mary,peter) + hates-(martha,peter)
\end{verbatim}
However, this example gives rise to some discussion of assumption
grammars. First of all, if the sentence preceding ``\texttt{and}'' 
happens to be
complete, the asserted hypothesis will not be used here
but may interfere with the analysis of subsequent ellipses.
To avoid this, the CHR rule can be refined so that
it only asserts the hypothesis if there is need for one;
replacing the unconditional assertion by the following piece of
code will do.
\begin{verbatim}
     find_constraint(=-(ref_object,_),_)
     -> =+(ref_object,[Obj2]) ; true
\end{verbatim}
In general, we believe that a more detailed control of the scope
of asserted hypothesis is needed than what is feasible with
assertion grammars. The explicit string indices are useful for this
as well as the low-level primitives of CHR and it seems that
suggestions for any such ``high-level'' scoping mechanism
can be implemented.\footnote{A tree-like scoping provided by
implications goals~\cite{mig-meta-90,gabbay-reyle-84,miller-89} does 
not seem sufficient
as some hypotheses may be relevant from, say, position
10 to 25 and others from 20 to 30. In addition, some hypotheses
may be temporarily overruled by others.}

Another feature missing in assumption grammars is the selection
of a \textit{best} hypothesis for, say, resolving a pronoun,
e.g., considering distances in the string or which
hypothesis has been applied most often.
We can sketch a rule as follows; the constraint before the backslash
serves as a test that there are applicable hypotheses.
\begin{flushleft}
     \texttt{\ \ \ \ +(active\char95 individual,[\char95,Gender],N)}\\
     \texttt{\ \ \ \ \ \ \ \ \ \ \ \ \ \ \ \ \ \ \ \ \ \ \ \ \ \ \ \ \ 
     \char92\ pronoun(Gender,N1,N2) <=> N<N1 |}\\
     \texttt{\ \ \ \ \ \ }\textit{select the best hypothesis of form}\\
     \texttt{\ \ \ \ \ \ \ \ \ \ +(active\char95individual,[X,Gender],\char95)}\\
     \texttt{\ \ \ \ \ \ \ \ \ \ }\textit{according to some criteria and remove
     it from the constraint store}{\ttfamily ,}\\
     \texttt{\ \ \ \ \ \ np(X,Gender,N1,N2).}
\end{flushleft}
If more hypotheses are feasible, they can be tested on backtracking.
--- We leave it to competent linguists to design such mechanisms. What
we have
indicated here is that CHR seems to provide a framework for
implementing them.

\section{Abduction and integrity constraints}
\label{abduction-section}
As several
authors~\cite{ChristiansenContext99,charniak-mcdermott-85,gabbey-kempson-pitt-94,weighted-abduction}
have noticed, abduction is a useful way
to conceive and to implement aspects
of natural language interpretation.
A deductive approach, as exercised in section~\ref{AGs},
represents the meaning of a text by a complex structure
attached to the start symbol and which has been
synthesized from meanings of its constituents.
By abduction, the meaning appear
as assumptions generated in parallel with the syntactic analysis
as they are needed, and integrity constraints
can check dynamically that the collected set of hypothesis is
consistent.
Consider the following DCG rule.
\begin{flushleft}
     \texttt{\ \ \ \ $A$ --> $B_1$, $\ldots$, $B_n$, \char123$P$\char125.}
\end{flushleft}
Now the instance of a nonterminal should be understood as the
presence of a \textit{good} phrase, e.g., a true sentence
or a noun phrase referring to an existing object,
and the related instance of $P$ is the precondition for the
composition of good subphrases itself to be a good phrase.
When formulating language interpretation as abduction, the unknown
is a body of assumptions that entails the set of all
such $P$ instances that have been applied.
The literature is rich of abduction procedures which can
do this job, 
e.g.,~\cite{Christiansen98,decker-96,denecker-deschreye98,kkt-abduction-98}.

In~\cite{AbdChr2000} we have shown how abduction with
integrity constraints can be expressed
in CHR. In the following we illustrate how to integrate it with
a CHR parser, however, without caring about the formal relation.
A DCG rule as above is written as the following CHR grammar rule:
\begin{flushleft}
     \texttt{\ \ \ \ $B_1$(N0,N1), $\ldots$, $B_n$(N$n$$-$$1$,N$n$) <=> 
$P$, $A$(N0,N$n$).}
\end{flushleft}
Applying it will introduce into the
constraint store the instance of \texttt{P}
that was necessary for the previous DCG rule to apply.\footnote{This
is likely to produce several instances of the same hypothesis but CHR
has means to avoid this.}
The \verb!active_individual(!$\cdots$\verb!)!
hypotheses generated in the example of section~\ref{AGs}
can be seen as an part of abductive explanation
why the text could have been sensibly and correctly uttered.
Gender of an individual may be recognized from its proper name in a
lexicon or
deduced from context. Thus the following
integrity constraint is relevant:
\begin{verbatim}
     active_individual(X,masc), active_individual(X,fem) <=> fail.
\end{verbatim}
Sentence meanings can be removed from the nonterminal and instead
be launched into the constraint store:
\begin{verbatim}
     np(Sub,_,N1,N2), verb(V,N2,N3), np(Obj,_,N3,N4) <=>
        fact(V,Sub,Obj), sent(N1,N4).
\end{verbatim}
Different anaphoric resolutions can be thinned out along the way by means of
``semantic'' integrity constraints such as:
\begin{verbatim}
     fact(likes,X,Y), fact(hates,X,Y) <=> fail.
     fact(loves,X,Y), fact(hates,X,Y) <=> fail.
     fact(hates,X,X) <=> fail.
\end{verbatim}
This will provide a correct interpretation of tricky
combinations such as ``Mary likes Martha. She hates her.''

Abduction provides also a way to work with negative hypotheses
by means of explicit negation as expressed in the
sketch of an integrity constraint,\break
% \begin{flushleft}
%      \texttt{\ \ \ \ not $H$, $H$ <=> fail.}
\texttt{not $H$, $H$ <=> fail.}
% \end{flushleft}
It means that anything can be asserted until the opposite has been
proved.

\medskip\noindent
We have indicated that CHR as grammar formalism seems capable
of integrating abductive text interpretation which has
several benefits over explicit synthesis of meanings.
However, the principle we have illustrated still needs
to be refined and its formal grounds to be established.

As we have described it, grammar rules have to be represented
as simplification rules, otherwise different hypothesis sets
for different parse trees will be mixed up. Thus alternatives
can only be explored by backtracking in the rule bodies.
However, it seems to be a matter of proper engineering to find ways to manage
sets of alternative and mutually exclusive hypotheses
in the same constraint store;
weighted abduction~\cite{weighted-abduction} seems to fit into such a 
model or a more structured approach based on fuzzy logic
could be used.
Integrity constraints should also behave in a less rude
way, by simple having wrong hypotheses to vanish instead
of provoking failure and thus backtracking.

\section{Conclusion and perspectives}
\label{conclusion-section}
What we have called CHR grammars relates to CHR the same way
that DCGs relate to Prolog. Grammar rules are written in
a systematic way as rules of the host
language and can be used directly as a parser
that inherits the computational characteristics
of the underlying machinery.
Where DCGs work top-down with backtracking,
CHR grammars works bottom-up.
In addition to provide an evaluation strategy
better suited for parsing, especially of
natural language, the CHR paradigm supports
the use of dynamically generated hypotheses,
e.g., for abduction or as in Assumption Grammars, that
is extremely useful for natural language analysis.

The experiences we have reported here are indeed promising
for the use of CHR as a grammatical formalism.
It provides a
high degree of flexibility and expressive power
for language processing
of which
we believe to have exposed only a small portion.

The comparison with DCGs may suggest to add a layer
of syntactic sugar to suppress arguments for syntactic indices
and to introduce high-level devices to replace the different
uses of them.
Take the simplification rule for an arithmetic plus
shown in section~\ref{complexity-section}.
By means of a suitable preprocessor, we may suggest to
write it in the following nice way.
\begin{verbatim}
     exp, [+], exp /- ([+];[')'];[eof]) <-> exp.
\end{verbatim}
We prefer to keep the order of lhs and rhs of the rules
to indicate their bottom-up nature and to
let the choice of propagation and simplification rule
be visible using symbols ``\texttt{-->}'' and ``\texttt{<->}''.
As in DCG, curly brackets can be used to indicated code
to be bypassed by the preprocessor, and notice that it is
relevant to apply this at both sides of a CHR grammar rule.
The ``\verb!/-!'' marker indicates a reference to the right context
of the sequence being reduced; an analogous ``\verb!-\!'' marker can be
used at the beginning of a rule
to indicate references to the left context.
To add \texttt{passive} pragmas to all but the leftmost grammar symbol
at the lhs as described in section~\ref{complexity-section},
a rule may be prefixed with an operator ``\texttt{ruleLR}'' or
a global directive ``\texttt{:- modeLR.}'' can affect all rules.
The LR mode should not be forced as it may affect the final result
for CHR grammars using reduction rules.

Our investigation has also exposed interesting topics for future
work: To study the possible relation between abduction
and assumption grammars (perhaps as instances of a more general
framework of hypothetical reasoning) and to design high-level facilities
for controlling the scope of hypotheses.
As we pointed out in section~\ref{abduction-section}, technical
work is needed in order to achieve a full bottom-up
and non-backtracking version of abduction.
Plans for the near future include
extraction of noun phrases to be used in a Danish
ontology-based search systems which is developed in parallel
with an  implementation using
traditional language processing tools.
A grammar for a substantial subset of Danish is under consideration
with contextual  and perhaps some kind of semantic analysis.

\bigskip\noindent
{\bf Acknowledgment:}
This research is supported in part by the OntoQuery
funded by the Danish Research Councils, and
the IT-University of Copenhagen.

\end{document}